\documentclass[12pt]{article}
\usepackage{times}
\usepackage{indentfirst,psfig}
\usepackage{epsfig}
\usepackage{cite}
\newcommand{\bey}[1]{\begin{eqnarray} \label{#1}}
\newcommand{\eey}{\end{eqnarray}}
\newcommand{\beq}[1]{\begin{equation} \label{#1}}
\newcommand{\eeq}{\end{equation}}
\makeatletter
\def\@evenhead{\vbox{\hbox to\hsize{\sl \headmsg \hfill}}}
\def\@oddhead{\vbox{\hbox to\hsize{\sl \headmsg \hfill }}}
\makeatother
\def\headmsg{}

\title{Evolvability is a selectable trait}

\author{David J. Earl and Michael W. Deem\
~\\
\hbox{}Department of Bioengineering and Department of Physics 
and Astronomy,\\
Rice University, Houston, Texas 77005-1892}

\begin{document}
\maketitle
\renewcommand{\baselinestretch}{1.3}
\tiny
\normalsize

{\flushleft
Corresponding author: Michael W. Deem, Department of Bioengineering 
and Department of Physics and Astronomy MS 142, Rice University,
6100 Main Street, Houston, Texas
77005--1892, 713-348-5852, 713-348-5811 (fax), 
mwdeem@chinook.rice.edu.
\bigskip

Manuscript information: 17 pages, 4 figures, 0 tables.
\bigskip

Word count: 112 words in abstract, 34507 characters in manuscript.
\emph{Proc.\ Natl.\ Acad.\ Sci.\ USA}
\bigskip

Physical Sciences. Physics.

Biological Sciences. Evolution.
}

\newpage
\def\headmsg{D. J. Earl and M. W. Deem, ``Evolvability 
is a selectable trait''}

\centerline{\textbf{ABSTRACT}}
\begin{quote}
Concomitant with the evolution of biological diversity
must have been the evolution of mechanisms that facilitate evolution,
due to the essentially infinite complexity of protein sequence space.
We describe how evolvability 
can be an object of Darwinian selection, emphasizing the
collective nature of the process.
We quantify our theory with computer
simulations of protein evolution.
These simulations demonstrate that
rapid or dramatic environmental change leads to selection for
greater evolvability.
The selective pressure for large scale genetic moves, such as DNA exchange,
becomes increasingly strong as the environmental conditions become
more uncertain.  Our results demonstrate that evolvability is a selectable
trait and allow for the explanation of a large body of experimental results.
\end{quote}

\newpage

\begin{quote}
Darwin was obsessed with variation. His books, considered
as an ensemble, devote much more attention to variation than
to natural selection, for he knew that no satisfactory theory
of evolutionary change could be constructed until the causes
of variation and the empirical rule of its form and amount
had been elucidated \cite{Gouldbook}.
\end{quote}

Whether or not the propensity to evolve, or evolvability 
\cite{Kirschner,Dawkins2,Radman}, can be an object of
Darwinian natural selection is a topic of great interest.
Causality would suggest not due to the apparently anticipatory nature
of evolvability \cite{new6,new2}.
Many within the field of evolutionary
biology are uncomfortable with the concept that evolvability 
is a selectable trait.
 A growing body of experimental data, however,
would be explained if evolvability were a selectable trait
\cite{Kidwell,Shapiro2,recomb1,Fedoroff,Shapiro3,Shapiro4,codon3,codon4,new5}. 

Organisms cannot evolve, 
or adapt, by germ line mutation to an environmental change within
their own lifetime. Does this mean that lineages
and individuals cannot be under selection for evolvability?  
While viability is the selection criterion,
the genotype that determines the viability arises in a mutated, evolved 
way from that of the previous generation as a result
of base substitution, recombination, transposition, and horizontal 
gene transfer.
These mutational
processes are the driving forces of evolution, and their rates fundamentally
determine evolvability. The perspective we here offer is that the
evolvability of an organism is defined by the rates of genetic
change, that genetic change is not always deleterious,
and that these rates of genetic change are not fixed and are
under selective pressure.  That is, the mechanisms that define
the rates of change are encoded in the genotype, and so they are
selectable.  An analogy with thermodynamics illuminates the issue:
How is free energy minimized in a physical system of particles,
despite the difficulty in defining the entropy of
a given configuration of the particles?   An ensemble of
particle configurations allows the definition of free energy and
the approach to thermodynamic equilibrium, just as
a population of evolving organisms allows the definition of and
selection for evolvability.

Within the framework of point mutation, base substitution, and
recombination, correlations of adaptation with function
have been observed.
It is known that immunoglobins have evolved
so that the mutation rates in complementary 
determining regions, where mutation is most likely to generate useful
variants, are much higher than those in framework regions 
\cite{codon1,codon4}. Recent data point to
a role for DNA polymerases in regulating the somatic hypermutation rate of 
immunoglobin genes \cite{codon2,codon3}.
Similarly, codon usage within the influenza hemagglutinin
protein appears to be
biased to favor more rapid antigenic drift \cite{codon4}.
Furthermore, in HIV-1 protease
the probability of mutation is not randomly distributed within the structure,
but rather concentrated at sites that alter the geometry of the protein
binding domain, conferring significant propensity for antigenic drift
\cite{Freire}.  
Such behavior is not mere curiosity and 
has widespread implications for drug design and the evolution of 
drug resistance \cite{Perelson}.
Stressful conditions may generally provoke
activation of error-prone polymerases,
triggering a large increase in adaptive rates
\cite{Rosenberg}. 
Not only point mutation, but also
recombination is widely appreciated to confer increased
evolvability \cite{recomb4,recomb1,recomb3}.
Recombination among the hemagglutinin and neuraminidase
proteins, for example, is believed to be a significant mechanism leading to the
emergence of new virulent strains of influenza
\cite{Frank}.
Computational and theoretical studies have also shown the 
persistence under selection of evolvability-enhancing moves in the
context of point mutation and recombination evolutionary dynamics
\cite{Wagner,Lenski,deBlasio,Travis,network1,network2}.

The selective forces that lead to the evolution and maintenance of
mechanisms for
rearrangement, deletion, transfer, and transposition of genetic material,
in as much as they lead to even greater evolution than point mutation
and recombination alone, are of great interest.
For example,
the development of antibiotic resistance in bacteria has evolved mainly
through the swapping of DNA pieces between the evolving bacteria 
\cite{Shapiro,Shapiro2}. 
Similarly,
the evolution of \emph{Eschericia coli} from \emph{Salmonella} is
thought to have occurred exclusively from DNA swapping \cite{len12}.
It has been proposed that the success of bacteria as a group
stems from a capacity to acquire 
genes from a large and diverse range of species \cite{recomb2}. 
It would appear, then, that large genetic moves are pervasive and
crucial to
evolutionary dynamics
\cite{len13,len12,len28,Shapiro,Shapiro2,Shapiro3,Shapiro4,len29,Fedoroff,Duret,Lonnig,recomb5,new1,new2,new5}. 
Concomitantly, evolvability is enhanced by these larger moves, as
shown experimentally for the case of DNA shuffling
\cite{len7,len8,len9,len10,recomb2,Lutz}. A key
question is whether selection for evolvability fosters the husbandry of
these moves.

We here address, from a theoretical point of view,
selection of evolvability in the presence of large-scale
genetic moves.
Although the use of the term evolvability has only recently come
to vogue in the scientific community, 
investigations into the evolution of adaptation 
go back several decades \cite{Clarke,Dawkins,Gould}. Prominent from
a theoretical perspective are works in population genetics 
\cite{Gillespie,Frank2}
and game theory \cite{MaynardSmith,MaynardSmith2,Sasaki}. Despite
the insights that these studies give as to the origin and maintenance 
of evolvability, evolution of and selection for
evolvability remains a hotly contested issue,
primarily due to the causality principle \cite{new6,new2}.
We here show that evolvability is selected for,
notwithstanding the constraints imposed by causality, when
a system is subject to a constant, random environmental change.
This selection of evolvability
occurs even when viability as a function of genotype is
an extremely complex function, with exponentially many optima, and when
the evolving system is unable to reach the global optimum of viability
in any one instance of the environment.
We demonstrate our results using computer simulations of protein
molecular evolution that incorporate selection in a 
varying environment. The genotype of a protein molecule is mapped to 
a complex phenotype using a random energy based model, in which all
assumptions and relevant parameters are known.  The selective pressure
for evolvability is shown to be greater for larger rates of environmental
change.  Interestingly, a generalized susceptibility of the system
correlates with the fluctuations in the environment, albeit not as a 
result of generalized linear response theory \cite{new7}.
The addition of selection for evolvability
as a phenomenological law to the toolbox of
evolutionary theory allows for the explanation of a large body of 
experimental results.

{\bf The Generalized Block $NK$ Model}.
Whether evolvability is selectable
has been a rather difficult question to answer, primarily because
observations in evolutionary biology
tend to be correlative in nature and
difficult upon which to make mechanistic conclusions.
Therefore, we here consider the dynamics of evolvability in
a well-defined theoretical model of protein evolution \cite{Bogarad_Deem}.
Within this model of protein structure and function, 
we have a fixed population of proteins, which we take
to be 1000.  We have a microscopic selection criterion,
which we take to be the folding and binding of a protein to a
substrate.  And we have a means of inducing
constant, random environmental change.

We model the molecular evolution of protein systems using a 
generalization of the $NK$ \cite{len14,len15,len16} and block 
$NK$ \cite{len17} models that has been used previously to study 
protein molecular evolution strategies \cite{Bogarad_Deem} and
the immune system response to vaccination and disease\cite{Deem}. 
The model includes a population of sequences, upon which
selection acts, and in which occur genetic mutations.  The mutational
hierarchy includes both point mutations and large-scale swapping moves,
akin to transposition or translocation events.  While the model does
not include recombination, such inclusion is not expected to change the
results, as swapping can be viewed as a powerful form of recombination
\cite{Bogarad_Deem}.  For example, linkage effects are mitigated
even more rapidly by swapping in our model than they would be by recombination.
The selection 
for greater swapping rates in more rapidly changing environments
observed in our model
parallels results found
in studies of the evolution of sex, where 
adaptation and variation in a heterogeneous environment is 
well-researched \cite{sex_evo}.

In the generalized block 
$NK$ model, each individual evolving protein sequence has an energy 
that is determined by secondary structural subdomain energies, $U^{\rm sd}$, 
subdomain-subdomain interaction energies, $U^{\rm sd-sd}$, and chemical binding
energies, $U^{\rm c}$. This energy is used as the selection criteria in 
our studies and is given by
\beq{101}
 U = \sum_{i = 1}^M U_{\alpha_i}^\mathrm{sd} +
\sum_{i > j = 1}^M U_{ i j }^\mathrm{sd-sd}
+ \sum_{i = 1}^P U_i^\mathrm{c} \ .
\eeq
Within our generalized block $NK$ model, each protein molecule is composed 
of $M=10$ secondary structural subdomains of $N=10$ amino acids in length.
We consider five chemically distinct amino acid classes (negative,
positive, polar, hydrophobic and other) and 
$L=5$ different types of subdomains 
(helices, strands, loops, turns and others).
We therefore have $L$ different subdomain energy functions of the $NK$ form
\beq{102}
U_{\alpha_{i}}^\mathrm{sd} = \frac{1}{\left[ M(N-K)\right]^{1/2}}
\sum_{j = 1}^{N-K+1} \sigma_{\alpha_{i}} \left(
a_j, a_{j+1}, \ldots, a_{j+K-1}
\right) \ ,
\eeq
where $a_{j}$ is the amino acid type of the $j$th amino acid in the
subdomain, and $\alpha_i$ is the type of the $i$th subdomain.
 As in previous studies, we consider the case where the 
range of the interactions within a subdomain is specified by $K=4$
 \cite{Bogarad_Deem,Deem}.
Here $\sigma_{\alpha_{i}}$ is a quenched Gaussian random number with zero
mean and a variance of unity, and it is different for each value of its 
argument for each of the $L$ subdomain types, $\alpha_{i}$. The 
interaction energy between secondary subdomain structures is given by
\bey{103}
U_{i j }^\mathrm{sd-sd} &=& \left[
\frac{2}{D M (M-1)} \right]^{1/2} \nonumber \\
&&\times 
\sum_{k=1}^D \sigma_{i j }^{(k)}
\left(
a_{j_1}^{(i)} , \dots,
a_{j_{K/2}}^{(i)};
a_{j_{K/2+1}}^{(j)}, \ldots
a_{j_{K}}^{(j)}
 \right) \ ,
\eey
where we consider $D=6$ interactions between secondary structures
 \cite{Bogarad_Deem,Deem}.
The zero-mean, unit-variance Gaussian $\sigma_{ i j }^{(k)}$ and 
the interacting amino acids,
$j_{1}, \ldots, j_{K}$, are selected at random for each interaction
($i, j, k $). In our model, $P=5$ amino acids contribute directly
to a binding event, as in a typical pharmacophore,
where the chemical binding energy of each amino acid
is given by
\beq{104}
U_i^\mathrm{c} = \frac{1}{\sqrt P} \sigma_i \left( a_i \right) \ ,
\eeq
where the zero-mean, unit-variance Gaussian
$\sigma_i$ and the contributing amino acid, $i$, are chosen at random.

{\bf System Evolution and Environmental Change}.
Our model system maintains a constant population of 1000 proteins, 
each protein of 100 amino acids in length and initially distinct
in sequence. 
The system evolves through the base substitution of single amino acids 
and via DNA swapping of amino acid subdomains
from structural pools representing the five different subdomain types, 
each containing 250 low-energy subdomain sequences. 
These moves represent the small-scale adaptation and the
large-scale, dramatic evolution that occur in Nature.
For protein $i$, $n_{\rm mut}(i)$ point mutations occur per sequence per round 
of selection.  In addition, for protein $i$, 
subdomain sequences are randomly replaced with
sequences from the same-type low-energy pools
with probability $p_{\rm swap}(i)$.

Following pool swapping and point mutations, selection occurs,
and the 20\% lowest-energy protein sequences are kept and amplified
to form the population of 1000 proteins for the next round 
of selection. The parameters
$p_{\rm swap}(i)$ and $n_{\rm mut}(i)$ are allowed to take a 
log-Gaussian random walk for each protein sequence.
This process is repeated for $N_{\rm gen}$ rounds of 
selection, after which an environmental change is imposed on the system 
with a severity that is characterized by the parameter, $p$
\cite{Deem}.
The parameter $p$ is the 
probability of (i) changing the type of each of the 10 subdomains in the 
protein sequences, $\alpha_i$ in Eq.\ \ref{102}, 
(ii) changing the amino acids and energies that are involved in 
subdomain-subdomain interactions, $j_k$ and $\sigma_{i j}^{(k)} $
in Eq.\ \ref{103}, and (iii) changing the amino acids and
energies that are involved in 
the chemical binding, $i$ and $\sigma_i$ in Eq.\ \ref{104}. We 
repeat the process for a total of 100 environmental changes and average 
our results over 1000 instances of the ensemble.
For each system studied, a
steady-state in $n_{\rm mut}$, $p_{\rm swap}$,
 and the average energies at the beginning, $\langle U \rangle _{\rm start}$,
and end, 
$ \langle U \rangle _{\rm end}$, of the dynamics in a single instance of
the environment is reached after fewer than 80 environmental 
changes in all cases.  We average the data over the last 20 environmental
changes. We study how the frequency of environmental change, 
$1/N_{\rm gen}$, and the severity of environmental change, $p$, affect the 
evolvability
of the protein sequences. A schematic diagram showing the molecular 
evolution of our protein system can be seen in Fig. 1. 

{\bf Selection For Evolvability}.
Shown in Fig. 2 are the steady state values of $p_{\rm swap}$ and 
$n_{\rm mut}$ that our protein system selects as a function of
imposed
frequency of environmental change, $1/N_{\rm gen}$, and severity
of environmental change, $p$. 
The DNA swapping moves that we propose have a high capacity for 
evolutionary change, as a significant number of amino acids may be altered in
a protein sequence in one swap move. It is clear that our systems
select for higher probabilities 
of DNA swapping moves, and thus evolvability, as the frequency and 
severity of environmental change increases. We stress the importance of this 
result. Mainstream evolutionary theory does not recognize a need for the 
selection of evolvability. More generally, we see that 
only in the limit of little or no environmental change,
$p_{\rm swap} \rightarrow 0$, do large scale changes tend to be
disfavored. 
The role of base substitution in our evolving system
is more complex. For more 
severe environmental changes and for higher frequencies of environmental 
change, the system depends more on DNA swapping than on
point mutation
to produce low-energy proteins. In these cases, since
the protein must make large changes to its sequence to adapt to the 
environmental change, selection results in high values of $p_{\rm swap}$, 
with base substitution having only a small effect on the energy of the 
protein. For less severe environmental changes and lower frequencies
of environmental change, base substitution is sufficient to
achieve the small modifications in protein sequence that are required
for adaptation to the environmental change.
Thus, we observe the higher dependence on 
$n_{\rm mut}$ and lower dependence on $p_{\rm swap}$ for small $p$.
In addition, as $1/N_{\rm gen} \rightarrow 0$, $n_{\rm mut} \rightarrow 0$
since
mutations tend to be deleterious in stable systems with no environmental
fluctuations.

Evolvability is intimately related to the diversity of a population.
At short times, evolvability can be quantified by the diffusion coefficient in protein
sequence space, $D_0$, which is given by the combined diffusion  
due to swapping of the subdomains and the point mutation of individual
amino acids \cite{Chandrasekhar}:
\beq{Diff}
D_0 = ({\rm const}) (10^2 \times 10 p_{\rm swap} + 1^2 n_{\rm mut}) \ .
\eeq
The overwhelming contribution to $D_0$ comes from the swapping step,
as the swapping move far more dramatically changes the sequence.
The short-time diffusion rate selected for reflects,
as a function of environmental change,
a balance between staying within a favorable basin of attraction, or niche,
and adaptation to a newly-created, superior niche.
As Fig.\ 2 shows, greater environmental change favors greater local
diffusion, as indicated by the monotonic increase of $p_{\rm swap}$ with $p$.

It is useful to regard base substitution as a means of fine tuning the 
protein sequences, whereas DNA swapping can be considered a
source of more substantial evolutionary change. This hierarchy within the
space of evolutionary moves becomes more 
apparent when studying the difference between starting and ending protein
sequences within one environment, as a function of 
$p$, $p_{\rm swap}$, and $n_{\rm mut}$.
The distance between protein sequences is characterized by the
Hamming distance between the respective amino acid sequences.
For a given $p$, the Hamming distance 
decreases only slightly as the frequency of environmental
change, $1/N_{\rm gen}$, increases, but has a very strong dependence on 
the severity of the environmental change, $p$, as shown in Fig.\ 3a.
The sensitivity of the Hamming distance also shows markedly different behavior
to $p_{\rm swap}$ and $n_{\rm mut}$, as shown in Fig.\ 3b.
 For state points with fixed $n_{\rm mut}$, 
$1/N_{\rm gen}$, and $p$,
the Hamming distance is strongly dependent on the value 
of $p_{\rm swap}$. However, for state points with fixed $p_{\rm swap}$,
$1/N_{\rm gen}$, and $p$,
the Hamming distance displays little or no variation with $n_{\rm mut}$.
The Hamming distance is a long-time measure of the evolvability of the system.
The long-time diffusion coefficient can be defined as the square 
of the Hamming distance
multiplied by the frequency of environmental change.
As Fig.\ 3a implies, the long-time evolvability, as measured by the long-time diffusion
coefficient, increases with both the severity and frequency of environmental change.

Due to the roughness of viability as a function of sequence,
the exploration performed by any particular individual is limited to a 
local basin of attraction defined by the short-time mutation rates, and so
more independent traces through sequence space allows for more thorough evolution.
In other words, the more diverse the starting population of individuals,
the greater potential there is for evolution.
Figure 3c shows the average variance of the energy values 
at the end of the dynamics within a single instance of the
environment as
a function of the severity and frequency of environmental change.
It is clear that the diversity increases monotonically with $p$ and $1/N_{\rm gen}$.

As we have seen, evolvability is quantifiable at any point in time via
measurement of diversity and the local mutation rates.  
For this reason, causality does not prevent selection for evolvability.
Because evolvability is an observable property, it can be selected for.

{\bf Susceptibility}. 
A further measure of long-time evolvability is the response, or susceptibility, of the
system to environmental change.
In Fig.\ 4a we plot the average energy at the
start, $\langle U \rangle_{\rm start}$,
and end, $\langle U \rangle_{\rm end}$,
of the dynamics within a single instance of the
environment.  This quantity is shown 
as a function of the severity, $p$, and frequency,
$1/N_{\rm gen}$,
 of environmental change.
It is apparent that 
at low frequencies of environmental change, populations with greater
diversity and variation, which are more evolvable, have slightly lower values of
$\langle U \rangle_{\rm end}$. 
There is also a clear increasing trend in 
$\langle U \rangle_{\rm start}$ as a function of $p$,
 which is a feature 
of the random energy model.
Considering the ending energy of a protein molecule within one
instance of the environment to be roughly the sum of $n$ Gaussian
terms from the generalized $NK$ model
\beq{105}
U_{\rm end} = \frac{1}{\sqrt n}
\sum_{i = 1}^{n} x_{i} \ .
\eeq
The starting energy of this protein molecule after an environmental
change is given by
\beq{106}
U_{\rm start} = \frac{1}{\sqrt n} \sum_{i = 1}^{n} x_{i}' \ ,
\eeq
where
\beq{107}
x_{i}' = \left \{ \begin{array}{cl}
x_{i} & {\rm with} \ {\rm probability} \ (1 - p) \\
x_{i}'' & {\rm with} \ {\rm probability} \ p
\end{array} \right. ,
\eeq
and
where $x_{i}^{''}$ are random Gaussian variables with zero mean 
$( \langle x_{i}^{''} \rangle = 0)$ whereas 
$x_{i}$ are evolved variables that are better than random, and
typically negative. 
Thus, the average starting energy of this protein molecule is
\beq{108}
U_{\rm start} = \frac{1}{\sqrt n}
\sum_{i = 1}^{n} ( p x_{i}^{''} + 
(1 - p) x_{i} ) \ .
\eeq
So, averaging over the values in the new environment
\beq{109}
U_{\rm start} = \frac{1}{\sqrt n} \sum_{i = 1}^{n} (1 - p) x_{i}
 = (1 - p) U_{\rm end} \ ,
\eeq
or, averaging over many environmental changes
\beq{109a}
\langle U \rangle_{\rm end} - \langle U \rangle_{\rm start} =
p  \langle U \rangle_{\rm end}  \ .
\eeq

This average reduction in the energy is a 
measure of the susceptibility of a system,
$\langle \Delta U \rangle / N_{\rm gen} =
(\langle U \rangle_{\rm end} - 
\langle U \rangle_{\rm start}) / N_{\rm gen}$.
In Fig.\ 4b we plot the susceptibility
of our system as a function of the severity of environmental change, $p$. 
For a fixed frequency of environmental change, the susceptibility is 
a linear function of the severity of environmental change, as in
Eq.\ (\ref{109a}).  This simple analysis captures the essence of the dynamics
that occurs in the correlated, generalized $NK$ model.
Fig.\ 4c shows that the probability 
distribution of the susceptibility is Gaussian in shape.  Note also that
the variance of the susceptibility increases with $p$ in Fig.\ 4c, and so
the linearity of the susceptibility in Fig.\ 4b is not simply the
result of a generalized fluctuation-dissipation theorem.

{\bf Implications for Evolution}.
Our results have wide-ranging implications for evolutionary
theory. In our model system, populations of protein molecules that 
are subject to greater environmental change select for higher rates of 
evolvability.
 The selection criteria that we use is not 
a measure of evolvability in any way, yet the system selects for 
evolvability based on the implicit energetic benefits of adaptation 
to environmental change.  In addition, there is no reason to assume that
selection is optimal.  In fact, systems optimal for one environment
tend to have too 
little evolvability and tend to be selected against when faced with 
the inevitability of change.

Given our results, we propose that it is not mere chance that 
highly evolvable species tend to be found in
rapidly changing environments
or that an environmental crisis can trigger
an increase in the rate of the evolution of a species.
Indeed, selection for evolvability
allows for the explanation of many data:
the existence
of somatic hypermutation in the immune system \cite{codon1,codon2,
codon3,codon4}, the evolution of 
drug resistance in species of bacteria \cite{Shapiro,Shapiro2},
and the occurrence and success of transpositional events 
in bacterial evolution \cite{Fedoroff,len12,Duret}.
A recently studied example from mammals is the
San Nicolas Island fox, which is a highly endangered species and
the most monomorphic sexually reproducing animal known.
This species, however,
is found to have high levels of genetic variation within the
major histocompatibility complex loci \cite{fox} that allows
for increased pathogen resistance.

We believe that our results are of great relevance to the field of
vaccine and drug design. Currently, the design of new vaccines
and drugs is largely based on the assumption that
pathogens evolve by local space searching in response to 
therapeutic and immune selection. However, it is clear that we 
must anticipate the evolutionary potential of large DNA
swapping events in the development of viruses, parasites,
bacteria, and cancers if we are to engineer effective methods of 
treating them.  How evolvability correlates with treatment strategy,
and how to drive pathogens into regions of low evolvability,
where they are most easily eradicated, is of significant importance
to efforts for vaccine and drug engineering.  

Specific pathogenic
examples of evolvability include the 
emergence of new influenza strains by a novel
hemagglutinin neuraminidase recombination, followed by
antigenic drift to a highly infectious strain \cite{Frank},
emergence of many new HIV strains with the spread of the
disease from its site of origin in Africa \cite{hiv_origin1,hiv_origin2},
and the increased emergence of new infectious diseases associated
with modern, post-World War II travel \cite{wwII}. Further, a recent study of
the dynamics of HIV-1 recombination suggests that HIV-1 may have
evolved high recombination rates in order to foster rapid
diversification and further its survival \cite{hiv_evol}.

Note that evolvability is not simply the observation that
new strains occur, rather it is the underlying probability with
which new strains are created by genetic modification.  These
new strains may proliferate, and be observed. Or these new strains
may fail, and not be observed to an appreciable extent.
Fundamental study of evolvability, then, requires an appreciation of these
underlying rates of genetic change.  These underlying rates,
such as polymerase error rates, recombination rates, and
transposition rates, are what selection for increased
evolvability may modulate \cite{Tan}.
These underlying rates of change are inheritable and can be altered
by mutation.  Study of these rates of genetic change, deconvoluted
from observed rates of evolution which are these rates multiplied
by a probability of survival, is of fundamental interest.

Intriguingly, we find that at low frequencies of environmental change,
populations that are subject to more severe
environmental changes can produce lower-energy individuals 
than populations that are not subject to environmental changes,
Fig.\ 4a.  Thus, under some conditions, adaptability can provide
global benefits.  This finding can be contrasted to the simpler
expectation often found in evolutionary literature that specialists
are better than generalists
\cite{Elena_Lenski}. In 
experimental studies of \emph{Chlamydomonas}, generalists that were
evolved in alternating light and dark conditions were found to be better
than their ancestors in both light and dark conditions, but less good
than specialists that had evolved exclusively in one of the environmental
conditions \cite{Reboud}. Studies of the 
evolution of \emph{Escherichia coli}
in constant and alternating temperatures produced similar results
\cite{Bennett_Lenski,Leroi_Lenski}. The nature of the environmental change in 
these studies is not completely random, as in our model.  In addition,
the number of rounds of selected evolution under each 
environmental condition is perhaps better defined within our model.
These experiments do point
to possible tests of our theory. For a species that is capable of DNA
swapping evolutionary moves, a systematic study of competency
as a function of the frequency of a random environmental change would
be of interest.  We predict that under some conditions, 
certain frequencies of environmental change will produce better
individuals, after a given number of rounds of evolution and selection,
than would be produced by evolution in a constant
environment. Different severities of environmental change could also be
imposed by altering the change in environmental
variables such as temperature, food concentrations, light conditions,
and exposure to disease, between samples.
With regards to susceptibility, we would expect the rate of change of 
viability within an environment to be higher in systems with more
frequent and harsher environmental changes due to greater
evolvability.

{\bf Summary}.
Not only has life evolved, but life has evolved to evolve.
That is, correlations within protein structure have evolved,
and mechanisms to manipulate these correlations have evolved
in tandem. The rates at which the various events within the hierarchy
of evolutionary moves occur are not random or arbitrary, but are
selected by Darwinian evolution. Sensibly, rapid or extreme environmental
change leads to selection for greater evolvability.  This selection is
not forbidden by causality and is strongest on the largest scale
moves within the mutational hierarchy.

Many observations within evolutionary biology, heretofore considered
evolutionary happenstance or accidents, are explained by selection for
evolvability.  For example, the vertebrate immune system shows that the
variable environment of antigens has provided selective pressure for the use of
adaptable codons
and low-fidelity polymerases during somatic hypermutation.
This selective pressure for adaptable codons has further
resulted in the altered codon usage of the immune system genes.
A similar driving force for biased codon usage
as a result of productively high mutation rates
is observed in the hemagglutinin protein of influenza A.
 Selection for evolvability explains the prevalence of transposons
among bacteria and recombination among higher organisms.
We suggest that therapeutics also confer selective pressure on the
evolvability of pathogens, and this driving force to antigenic drift
should be considered in drug and vaccine design efforts.

This research is supported by the National Institutes of Health.
The authors thank Kevin R. Foster for a careful reading of the
manuscript.
\newpage

\bibliography{evolvability}

\begin{thebibliography}{10}
\newcommand{\enquote}[1]{``#1''}

\bibitem{Gouldbook}
Gould, S.~J. (1983) \textit{Hen's Teeth and Horse's Toes} (Norton, W. W. \&
  Company).

\bibitem{Kirschner}
Kirschner, M. \& Gerhart, J. (1998) \textit{Proc. Natl. Acad. Sci. USA}
  \textbf{95}, 8420--8427.

\bibitem{Dawkins2}
Dawkins, R. (1989) In \textit{Artificial Life}, ed. Langton, C.~G.
  (Addison-Wesley, New York), pp. 201--220.

\bibitem{Radman}
Radman, M., Matic, I. \& Taddei, F. (1999) \textit{Ann. N.Y. Acad. Sci.}
  \textbf{870}, 146--155.

\bibitem{new6}
Chicurel, M. (2001) \textit{Science} \textbf{292}, 1824--1827.

\bibitem{new2}
Partridge, L. \& Barton, N.~H. (2000) \textit{Nature} \textbf{407}, 457--458.

\bibitem{Kidwell}
Kidwell, M.~G. (1997) \textit{Proc. Natl. Acad. Sci. USA} \textbf{94},
  7704--7711.

\bibitem{Shapiro2}
Shapiro, J.~A. (1997) \textit{Trends in Genetics} \textbf{13}, 98--104.

\bibitem{recomb1}
Barton, N.~H. \& Charlesworth, B. (1998) \textit{Science} \textbf{281},
  1986--1990.

\bibitem{Fedoroff}
Fedoroff, N. (2000) \textit{Proc. Natl. Acad. Sci. USA} \textbf{97},
  7002--7007.

\bibitem{Shapiro3}
Shapiro, J.~A. (2002) \textit{J. Biol. Phys.} \textbf{28}, 745--764.

\bibitem{Shapiro4}
Shapiro, J.~A. (2002) \textit{Ann. N.Y. Acad. Sci.} \textbf{981}, 111--134.

\bibitem{codon3}
Storb, U. (2001) \textit{Nature Immunol.} \textbf{2}, 484--485.

\bibitem{codon4}
Plotkin, J.~B. \& Dushoff, J. (2003) \textit{Proc. Natl. Acad. Sci. USA}
  \textbf{100}, 7152--7157.

\bibitem{new5}
Caporale, L.~H. (2003) \textit{American Scientist} \textbf{91}, 234--241.

\bibitem{codon1}
Kepler, T.~B. (1997) \textit{Mol. Biol. Evol.} \textbf{14}, 637--643.

\bibitem{codon2}
Friedberg, E.~C., Feaver, W.~F. \& Gerlach, V.~L. (2000) \textit{Proc. Natl.
  Acad. Sci. USA} \textbf{97}, 5681--5683.

\bibitem{Freire}
Freire, E. (2002) \textit{Nature Biotech.} \textbf{20}, 15--16.

\bibitem{Perelson}
Kepler, T.~B. \& Perelson, A.~S. (1998) \textit{Proc. Natl. Acad. Sci. USA}
  \textbf{95}, 11514--11519.

\bibitem{Rosenberg}
Rosenberg, S.~M. (2001) \textit{Nature Rev. Genetics} \textbf{2}, 504--515.

\bibitem{recomb4}
Pepper, J.~W. (2003) \textit{BioSystems} \textbf{69}, 115--126.

\bibitem{recomb3}
Colegrave, N. (2002) \textit{Nature} \textbf{420}, 664--666.

\bibitem{Frank}
Frank, S.~A. (2002) \textit{Immunology and Evolution of Infectious Disease}
  (Princeton University Press, Princeton, NJ).

\bibitem{Wagner}
Wagner, G.~P. \& Altenberg, L. (1996) \textit{Evolution} \textbf{50}, 967--976.

\bibitem{Lenski}
Lenski, R.~E., Ofria, C., Pennock, R.~T. \& Adami, C. (2003) \textit{Nature}
  \textbf{423}, 139--144.

\bibitem{deBlasio}
Blasio, F. V.~D. (1999) \textit{Phys. Rev. E} \textbf{60}, 5912--5917.

\bibitem{Travis}
Travis, J. M.~J. \& Travis, E.~R. (2002) \textit{Proc. R. Soc. Lond. B}
  \textbf{269}, 591--597.

\bibitem{network1}
Siegal, M.~L. \& Bergman, A. (2002) \textit{Proc. Natl. Acad. Sci. USA}
  \textbf{99}, 10528--10532.

\bibitem{network2}
Bergman, A. \& Siegal, M.~L. (2003) \textit{Nature} \textbf{424}, 549--552.

\bibitem{Shapiro}
Shapiro, J.~A. (1992) \textit{Genetica} \textbf{86}, 99--111.

\bibitem{len12}
Lawrence, J.~G. (1997) \textit{Trends Microbiol.} \textbf{5}, 355--359.

\bibitem{recomb2}
Zhang, Y.-X., Perry, K., Vinci, V.~A., Powell, K., Stemmer, W. P.~C. \& del
  Cardayre, S.~B. (2002) \textit{Nature} \textbf{415}, 644--646.

\bibitem{len13}
Pennisi, E. (1998) \textit{Science} \textbf{281}, 1131--1134.

\bibitem{len28}
Gilbert, W. (1978) \textit{Nature} \textbf{271}, 501.

\bibitem{len29}
Gilbert, W., DeSouza, S.~J. \& Long, M. (1997) \textit{Proc. Natl. Acad. Sci.
  USA} \textbf{94}, 7698--7703.

\bibitem{Duret}
Duret, L., Marais, G. \& Biemont, C. (2000) \textit{Genetics} \textbf{156},
  1661--1669.

\bibitem{Lonnig}
Lonnig, W.-E. \& Saedler, H. (2002) \textit{Ann. Rev. Genet.} \textbf{36},
  389--410.

\bibitem{recomb5}
Levin, B.~R. \& Bergstrom, C.~T. (2000) \textit{Proc. Natl. Acad. Sci. USA}
  \textbf{97}, 6981--6985.

\bibitem{new1}
Doolittle, W.~F. (2000) \textit{Sci. Amer.} \textbf{282}, 90--95.

\bibitem{len7}
Stemmer, W. P.~C. (1994) \textit{Nature} \textbf{370}, 389--391.

\bibitem{len8}
Crameri, A., Raillard, S.~A., Bermudez, E. \& Stemmer, W. P.~C. (1998)
  \textit{Nature} \textbf{391}, 288--291.

\bibitem{len9}
Zhang, J.-H., Dawes, G. \& Stemmer, W. P.~C. (1997) \textit{Proc. Natl. Acad.
  Sci. USA} \textbf{94}, 4504--4509.

\bibitem{len10}
Moore, J.~C., Jin, H.-M., Kuchner, O. \& Arnold, F.~H. (1997) \textit{J. Mol.
  Evol.} \textbf{272}, 336--347.

\bibitem{Lutz}
Lutz, S. \& Benkovic, S.~J. (2000) \textit{Curr. Opin. Biotech.} \textbf{11},
  319--324.

\bibitem{Clarke}
Clarke, B.~C. (1979) \textit{Proc. Roy. Soc. Lond. B} \textbf{205}, 453--474.

\bibitem{Dawkins}
Dawkins, R. \& Krebs, J.~R. (1979) \textit{Proc. Roy. Soc. Lond. B}
  \textbf{205}, 489--512.

\bibitem{Gould}
Gould, S.~J. \& Lewontin, R.~C. (1979) \textit{Proc. Roy. Soc. Lond. B}
  \textbf{205}, 581--598.

\bibitem{Gillespie}
Gillespie, J.~H. (1991) \textit{The causes of molecular evolution} (Oxford
  University Press, Oxford).

\bibitem{Frank2}
Frank, S.~A. \& Slatkin, M. (1990) \textit{Am. Nat.} \textbf{136}, 244--260.

\bibitem{MaynardSmith}
Smith, J.~M. (1979) \textit{Proc. Roy. Soc. Lond. B} \textbf{205}, 475--488.

\bibitem{MaynardSmith2}
Smith, J.~M. (1982) \textit{Evolution and the theory of games} (Cambridge
  University Press, Cambridge).

\bibitem{Sasaki}
Sasaki, A. \& Ellner, S. (1995) \textit{Evolution} \textbf{49}, 337--350.

\bibitem{new7}
Sato, K., Ito, Y., Yomo, T. \& Kaneko, K. (2003) \textit{Proc. Natl. Acad. Sci.
  USA} \textbf{100}, 14086--14090.

\bibitem{Bogarad_Deem}
Bogarad, L.~D. \& Deem, M.~W. (1999) \textit{Proc. Natl. Acad. Sci. USA}
  \textbf{96}, 2591--2595.

\bibitem{len14}
Kauffman, S. \& Levin, S. (1987) \textit{J. Theor. Biol.} \textbf{128}, 11--45.

\bibitem{len15}
Kauffman, S.~A. (1993) \textit{The Origins of Order} (Oxford University Press,
  New York).

\bibitem{len16}
Kauffman, S.~A. \& MacReady, W.~G. (1995) \textit{J. Theor. Biol.}
  \textbf{173}, 427--440.

\bibitem{len17}
Perelson, A.~S. \& Macken, C.~A. (1995) \textit{Proc. Natl. Acad. Sci. USA}
  \textbf{92}, 9657--9661.

\bibitem{Deem}
Deem, M.~W. \& Lee, H.~Y. (2003) \textit{Phys. Rev. Lett.} \textbf{91}, 068101.

\bibitem{sex_evo}
Michod, R.~E. \& Levin, B.~R., eds. (1988) \textit{The Evolution of Sex}
  (Sinauer Associates Inc., Sunderland, MA).

\bibitem{Chandrasekhar}
Chandrasekhar, S. (1943) \textit{Rev. Mod. Phys.} \textbf{15}, 1--89.

\bibitem{fox}
Aguilar, A., Roemer, G., Debenham, S., Binns, M., Garcelon, D. \& Wayne, R.~K.
  (2004) \textit{Proc. Natl. Acad. Sci. USA} \textbf{101}, 3490--3494.

\bibitem{hiv_origin1}
Zhu, T.~F., Korber, B.~T., Nahmias, A.~J., Hooper, E., Sharp, P.~M. \& Ho,
  D.~D. (1998) \textit{Nature} \textbf{391}, 594--597.

\bibitem{hiv_origin2}
Gao, F., Bailes, E., Robertson, D.~L., Chen, Y.~L., Rodenburg, C.~M., Michael,
  S.~F., Cummins, L.~B., Arthur, L.~O., Peeters, M., Shaw, G.~M., Sharp, P.~M.
  \& Hahn, B.~H. (1999) \textit{Nature} \textbf{397}, 436--441.

\bibitem{wwII}
Lederberg, J., Shope, R.~E. \& S.~C.~Oaks, J., eds. (1992) \textit{Emerging
  Infections: Microbial Threats to Health in the United States} (National
  Academy Press, Washington, D.C.).

\bibitem{hiv_evol}
Levy, D.~N., Aldrovandi, G.~M., Kutsch, O. \& Shaw, G.~M. (2004) \textit{Proc.
  Natl. Acad. Sci. USA} \textbf{101}, 4204--4209.

\bibitem{Tan}
Tan, T., Bogarad, L.~D. \& Deem, M.~W. (2004) \textit{J. Mol. Evol.} To appear.

\bibitem{Elena_Lenski}
Elena, S.~F. \& Lenski, R.~E. (2003) \textit{Nature Rev. Genetics} \textbf{4},
  457--469.

\bibitem{Reboud}
Reboud, X. \& Bell, G. (1997) \textit{Heredity} \textbf{78}, 507--514.

\bibitem{Bennett_Lenski}
Bennett, A.~F. \& Lenski, R.~E. (1993) \textit{Evolution} \textbf{47}, 1--12.

\bibitem{Leroi_Lenski}
Leroi, A.~M., Lenski, R.~E. \& Bennett, A.~F. (1994) \textit{Evolution}
  \textbf{48}, 1222--1229.

\end{thebibliography}

\clearpage

\begin{figure}[t]
\caption{Schematic diagram showing the evolution of the protein system.
\hspace{3in}
\label{fig:schematic}}
\end{figure}

\begin{figure}[h]
\caption{$n_{\rm mut}$ (dashed line) and $p_{\rm swap}$ (solid line)
as a function 
of the frequency of environmental change, $1/N_{\rm gen}$, for different values 
of the severity of environmental change, $p$. The statistical errors in the
results are smaller than the symbols on the figure.
\label{fig:nmut_pswap}}
\end{figure}

\begin{figure}[h]
\caption{a) Hamming distance as a function of the severity of environmental
change, $p$, for the state points in Fig.~\ref{fig:nmut_pswap}.
b) Hamming distance as a function of $n_{\rm mut}$ (dashed line) and 
$p_{\rm swap}$ (solid line) for fixed $N_{\rm gen}=15$ and for different
severities of environmental change, $p$. In displaying the Hamming distance
dependence on $n_{\rm mut}$
($p_{\rm swap}$), we fix $p_{\rm swap}$ ($n_{\rm mut}$) 
to the selected values from Fig.~\ref{fig:nmut_pswap}. The selected values
of $n_{\rm mut}$ and $p_{\rm swap}$ at each state point are shown by 
light and dark circles respectively.
c) Average variance, $\sigma^2_{U_{\rm end}}$,
 of the energy of a population at the end of an evolution,
$U_{\rm end}$, as a function of the severity of environmental
change, $p$, for different frequencies of environmental change, $N_{\rm gen}$.
\label{fig:Hamming}}
\end{figure}

\begin{figure}[!h]
\caption{a) 
Average energy immediately after, $\langle U \rangle_{\rm start}$,
and immediately before, $\langle U \rangle_{\rm end}$, an environmental
change as a function of the severity of environmental change, $p$, for 
different frequencies of environmental change.
b) Average change in energy, $\langle \Delta U \rangle$, multiplied
by the frequency of environmental change, $1/N_{\rm gen}$,
 as a function of the severity
of environmental change, $p$.
c) Probability distribution of the susceptibility for different values 
of the severity of environmental change, $p$, for a fixed frequency of
environmental change, $1/N_{\rm gen} = 0.1$.
\label{fig:Uvp}}
\end{figure}

%\end{document}

\clearpage

\begin{center}
\epsfig{file=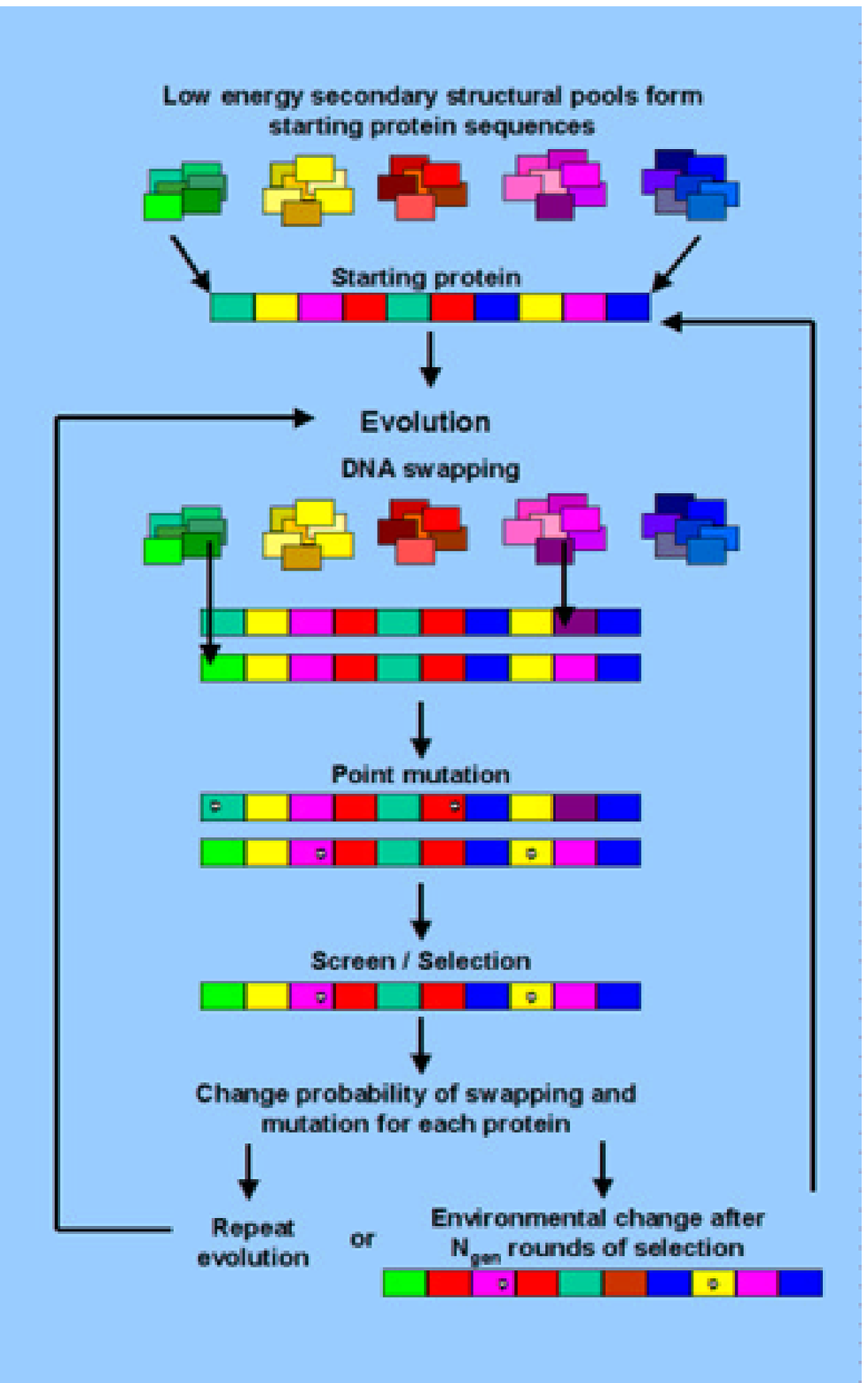,height=6.0in}
\end{center}
Figure~\ref{fig:schematic}.
Earl and Deem, ``Evolvability is a selectable trait''

\clearpage
\begin{center}
\epsfig{file=figure2.eps,height=3in,clip=,angle=0}
\end{center}
Figure~\ref{fig:nmut_pswap}.
Earl and Deem, ``Evolvability is a selectable trait''

\clearpage
\begin{center}
\epsfig{file=fig3a.eps,height=2.8in,clip=,angle=0}
\epsfig{file=fig3b.eps,height=2.8in,clip=,angle=0}
\epsfig{file=fig3c.eps,height=2.8in,clip=,angle=0}
\end{center}
Figure~\ref{fig:Hamming}.
Earl and Deem, ``Evolvability is a selectable trait''

\clearpage
\begin{center}
\epsfig{file=fig4a.eps,height=2.8in,clip=,angle=0}
\epsfig{file=fig4b.eps,height=2.8in,clip=,angle=0}
\epsfig{file=fig4c.eps,height=2.8in,clip=,angle=0}
\end{center}
Figure~\ref{fig:Uvp}.
Earl and Deem, ``Evolvability is a selectable trait''

\end{document}